\documentclass[twocolumn,showpacs,preprintnumbers,prb,amsmath,amssymb,floatfix]{revtex4}

\usepackage{graphicx}
\usepackage{epsfig}
\usepackage{dcolumn}
\usepackage{bm}

\newcommand{\bb}{\hspace{1mm}}
\newcommand{\vst}{$v^{*}$}
\newcommand{\tc}{$T_{C}$ }
\newcommand{\tcy}{$T_{C}$}
\newcommand{\teps}{$\tau_\epsilon\bb$}
\newcommand{\tepse}{\tau_\epsilon}
\newcommand{\rate}{$\tau_\epsilon^{-1}$}
\newcommand{\ybco}{Y$_{1}$Ba$_{2}$Cu$_{3}$O$_{7-\delta}$ }

\begin{document}

\preprint{Physical Review B {\bf 74}, 064512 (2006)}

\title{Energy Relaxation at a Hot-Electron Vortex Instability}

\author{James M. Knight}
 \email{knight@sc.edu}
\author{Milind N. Kunchur}%
 \email{kunchur@sc.edu}
 \homepage{http://www.physics.sc.edu/kunchur}
\affiliation{Department of Physics and Astronomy, University of South Carolina,
 Columbia, SC 29208}

\date{Recieved on June 28, 2006; published on August 24, 2006}

\begin{abstract}
At high dissipation levels, vortex motion in a superconducting
film has been observed to become unstable at
a certain critical vortex velocity \vst. At substrate temperatures 
substantially below \tcy, the observed
behavior can be accounted for by a model 
in which the electrons
reach an elevated temperature relative to the phonons and the substrate.
Here we examine 
the underlying assumptions concerning energy flow and relaxation times
in this model. A calculation of the rate of energy
transfer from the electron gas to the lattice finds that at the
instability, the electronic temperature reaches a very high value close
to the critical temperature. Our calculated energy relaxation times
are consistent with those deduced from the experiments. 
We also estimate the phonon mean free path and assess
its effect on the flow of energy in the film. 
\end{abstract}
\maketitle

\section{\label{sec:level1}Introduction}

When a film of a type II superconductor is placed in a magnetic 
field large enough to permit penetration of vortices, a transport current
in the film acts on the vortices through a Lorentz force that is opposed
by a pinning force and, eventually, by a drag force.
When the Lorentz force
exceeds the pinning force, the vortices are set into motion 
and the drag force comes into play. When the Lorentz force is substantially
larger than the pinning forces but the transport current is still small 
compared to the depairing current, previous experiments 
\cite{K1,K2,K2a} showed that the resulting dissipation is reasonably well
described by the Bardeen-Stephen (BS) model \cite{BS}. 
In this region it is Ohmic, but as the current is increased, 
it becomes non-linear and
eventually reaches an instability manifested by a discontinuous increase
in voltage. At 
temperatures not far below the critical temperature, the instability 
has been studied in a classic paper \cite{LOrev} by Larkin and Ovchinnikov (LO).
They showed that the electron distribution 
departs from a thermal distribution at high  
vortex velocities, changing the superconducting order parameter 
and altering the drag
force on the vortices. They predicted a 
non-linearity in the current-voltage 
characteristic and an instability in the vortex motion when
the vortices reach a critical velocity \vst. The LO instability is due to 
a decrease in the drag force with increasing vortex velocity, 
accompanied by a decrease in vortex size. LO showed that the 
critical velocity is independent of the magnetic field. Early experiments on
low-\tc systems \cite{loexpts} confirmed Larkin and Ovchinnikov's results and 
predictions. Subsequent experiments on \ybco (YBCO) 
by Doettinger, Huebener, Gerdemann, K\"{u}hle, Anders, Tr\"{a}uble, and Vill\`{e}gier \cite{heub} and by Xiao and Ziemann \cite{XZ},
also confirmed LO behavior. 

However, experiments carried out at lower temperatures \cite{K3,K4} on YBCO,
showed a non-linearity and instability with a very different dependence of 
\vst\bb on the magnetic field $B$.  Analysis \cite{K3,K4} showed that the new 
behavior could be accounted for by a simple model in
which the electron gas has a thermal-like distribution
function characterized by a higher temperature than the lattice and bath. 
Larkin and Ovchinnikov did, in fact, suggest this possibility
in their original paper \cite{LO} without exploring its consequences.
As the electron temperature rises, the resulting increase in resistivity
causes a decrease in current above a certain electric field and hence a
non-monotonic response. This model yields a critical 
vortex velocity \vst\bb at instability 
that is proportional to $1/\sqrt{B}$, as seen in the low-$T$ experiments.
Some of the essential consequences of such a hot-electron instability
were calculated in our earlier papers and shown to be consistent with
experimental observations. 

In the present work some of the
simplifying assumptions and restrictions in the previous calculations
have been removed and more complete calculations have been carried out:

1.) The rate \rate\bb of transfer of energy from the electron
gas to the lattice---which plays a crucial role in determining the electron 
temperature---was taken as a constant in previous discussions of 
the model. In this paper we show that it can be expected
to have a strong temperature dependence. This temperature dependence of
\teps is now included in our numerical calculations of the current-voltage
curves. We find that the general shape of the current-voltage relation
is not very sensitive to the temperature variation of \teps 
because the electron gas passes rapidly from the bath temperature
to a temperature not far below \tc\bb before any significant
non-linearity is manifested. This is a consequence of the very small 
low-temperature specific heat of a superconducting electron gas.
However, the strong temperature variation of the relaxation 
time gives a sensitive measure of the electron temperature.
Evaluation of \teps from the data near the instability point
indicates an electron temperature much higher than the bath temperature, 
supporting the heated electron picture of the instability. 
The calculation of this electron-lattice energy 
relaxation time is presented in Section III below. 

2.) In our previous work\cite{K3,K4}, we assumed the film thickness 
to be negligible compared to the phonon mean free path, so that the
phonon temperature is uniform throughout the film. 
In this work, we remove this assumption and take phonon lifetime effects
into account. The phonons will not necessarily be at the bath temperature,
and will have a non-thermal distribution which varies with position in the 
film. Phonon lifetime effects can be taken into account
following work by Bezuglij and Shklovsky \cite{BjSy}, who solved the
phonon kinetic equation for a thin film. 
The non-thermal phonon distribution found in this solution can be 
incorporated into our calculation of the energy transfer rate, and 
provides a correction to our earlier results. This result is derived 
in Section IV below. 

We begin in Section II by giving a description of the
model presenting some new numerical results for the
current-voltage curves under various conditions and for the critical
parameters at the instability.

\section{\label{sec:level2}Model for Instability}

The macroscopic fields in a type II superconductor carrying a 
transport current are related to the velocity of the vortices  
by the fundamental relation 
\begin{equation}
v=\frac{E}{B}c, \label{vEB}
\end{equation}
which follows from the law of induction.
This equation can be used to find the electric field once the vortex velocity 
is determined by considering the fundamental
dissipative processes in the medium. Elastic forces can be shown to be 
negligible.
One of the dissipative processes is the scattering of 
normal electrons in the vortex
core and quasiparticles outside the core first treated by
Bardeen and Stephen. They found that the transport current density $j$ is 
expressed in terms of the upper critical field $H_{c2}$ and the normal
resistivity $\rho_n$ by
\begin{equation}
j = \frac{H_{c2}}{\rho_n}\frac{E}{B}.
\label{j1}
\end{equation}
Later treatments \cite{LOrev,BS+,kopnin} 
take into account the relaxation of the order parameter during passage
of the vortex first treated by Tinkham \cite{Tink}. They give results which 
vary with circumstances, but agree with Eq.(\ref{j1})
to within a numerical factor of order 1.  

These energy dissipation mechanisms raise the energy of the electrons,
and this energy subsequently relaxes to the lattice. 
The assumption of our model is that 
the electron-electron scattering time is small enough compared to 
the electron-phonon inelastic scattering time that the electron gas
remains in internal thermal equilibrium at a temperature higher 
than the lattice temperature. 
The plausibility of the assumption can be checked by estimating the cross-over 
temperature below which electron-electron scattering is 
dominant. The standard estimates \cite{Abri} of the scattering rates
$\tau_{ee}^{-1} = \eta \epsilon_F/T^2$ and of
$\tau_{ep}^{-1} = \eta^3 \omega_D^2/T^3$ then give a cross-over temperature
of the order of 100 K for parameters appropriate to YBCO\cite{nist}. This
temperature is indeed higher than the range of interest in the experiments. 

Changes in the energy density of the electron gas can be described by
a rate equation that includes the work done by the electric field
and the exchange of energy with the lattice. If we assume that the
exchange can be described approximately by an energy relaxation time
\teps, then the equation is
\begin{equation}
\frac{du}{dt} = jE - \frac{u(T') - u(T_p)}{\tepse(T', T_p)}, 
\label{dudt}
\end{equation}
where \teps can depend on the phonon temperature $T_p$ 
as well as on the elevated electron temperature $T'$. 
We argue below that the dependence of \teps on 
$T_p$ is weak enough to be ignored in the relevant range of 
temperatures and the relevant energy transfer rates between the
lattice and the bath. The quasiparticles transfer the energy they receive
from the transport current to the lattice at a rate much higher than
it is radiated back, and the energy then flows from the lattice to the bath.
Thus \teps can be assumed to depend only on $T'$, and 
we can write the steady-state equation
\begin{equation}
jE\tepse = \int_{T_p}^{T'}c(T)dT, \label{jet}
\end{equation}
where the energy difference in Eq.(\ref{dudt}) has been expressed in terms
of the electronic specific heat per unit volume.

Equations (\ref{j1}) - (\ref{jet})
determine the relationship between the electric field, the current
density and the temperature. 
The temperature dependence of the specific heat and the upper
critical field are taken from standard BCS theory \cite{TWHH}.
In calculating the specific heat, the temperature dependence of the 
gap was taken from BCS theory and its magnitude was multiplied by
a factor to give
the observed zero-temperature gap \cite{D00} and critical temperature.
In the next section, we calculate the energy relaxation time and its 
temperature dependence. 

\begin{figure}[ht]
\epsfxsize=80mm
\epsffile{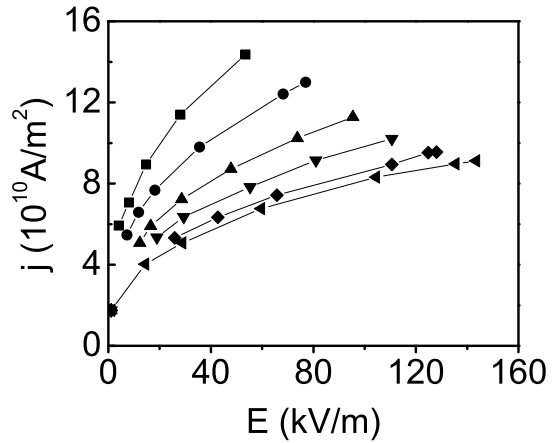}
\caption{Experimental curves of current density vs electric field 
in YBCO for flux density (beginning with the upper curve) 
$B =$ 3, 5, 8, 11, 14, and 16 Tesla.}
\label{expdat}
\end{figure}

Typical results of the model are presented in the following figures. Fig.
\ref{FJEB} shows the calculated current-voltage curves for different magnetic
fields. The shape of the curves is in general qualitative agreement 
with the experimental data shown in Fig \ref{expdat}.  
The strength parameter $b$ of the electron-phonon
coupling, defined in the next section, was adjusted to obtain agreement 
with the values of $E$ and $j$ at the peak.  We comment on the choice of 
$b$ in the next section.
 
\begin{figure}[hb]
\epsfxsize=80mm
\epsffile{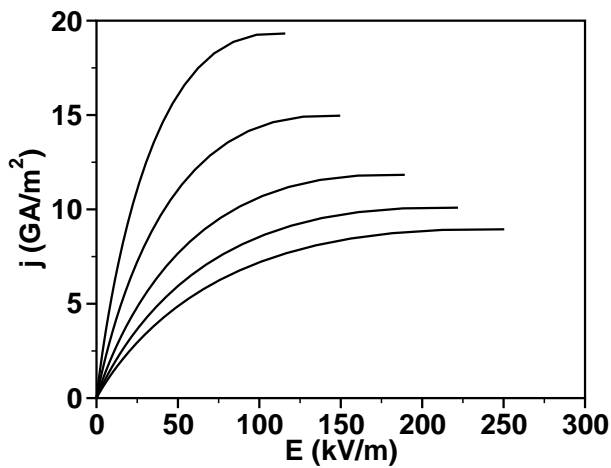}
\caption{Effect of increasing flux density on the current density vs
electric field curve 
calculated in the model with variable \teps. Values
of $B$ beginning at the upper curve are 3, 5, 8, 11, and 14 Tesla.}
\label{FJEB}
\end{figure} 
\begin{figure}[hb]
\epsfxsize=80mm
\epsffile{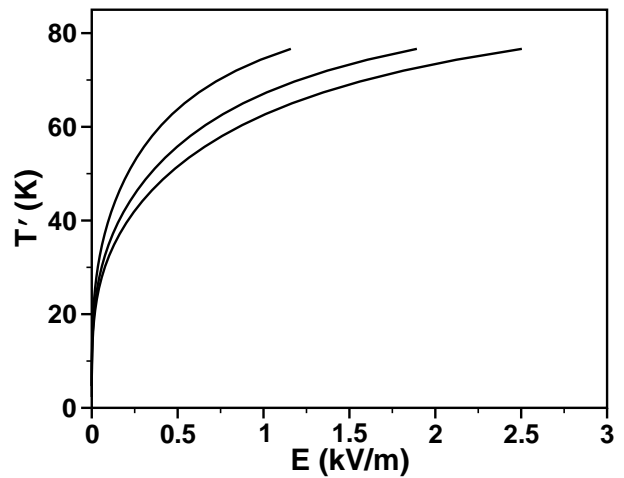} 
\caption{Calculated electron temperature $T'$ vs electric field $E$ 
for $B = 3, 8, {\rm and} 14$ Tesla, with $B$ increasing from left to 
right.}
\label{FTEB}
\end{figure} 

\begin{figure}[hb]
\epsfxsize=80mm
\epsffile{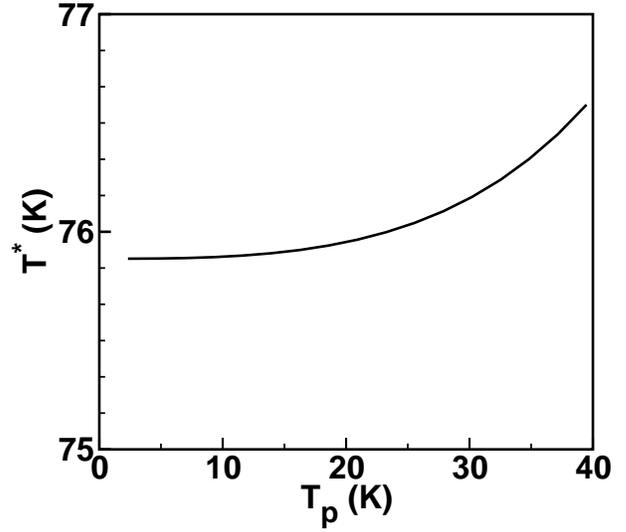} 
\caption{Electron temperature at the instability vs. phonon temperature}
\label{FTT0}
\end{figure}

The onset of the unstable region in the current-voltage response does not
require explicitly invoking the forces on the vortices
in treatment of the model. Rather, the instability 
appears in the result as a region of negative differential
conductivity, where $j$ decreases as a function of $E$. The region begins
at the value $E^{*}$ of the field that can be determined by calculating
$dj/dE$ from Eq.(\ref{jet}), setting the result equal to zero, 
and solving for $E$:
\begin{equation}
E^{*} = \sqrt{\frac{C\rho_n B}{H_{c2}\tepse^{'} + (H_{c2})^{'}\tepse)}}, \label{Estar}
\end{equation}
where primes indicate differentiation with respect to temperature.
\footnote{A secondary consequence of this 
instability is a fragmentation of the flux flow in the region of
negative $dj/dE$ resulting in steps in the current-voltage curves.
Such steps have been observed in constant-voltage
measurements at high dissipations \cite{steps,shear}.} 
The experimentally 
well-verified $\sqrt B$ dependence of $E^{*}$ 
follows provided the temperature $T^{*}$ at the instability 
is independent or weakly dependent on $B$ so that the 
temperature-dependent factors C, \teps, and $H_{c2}$ 
in Eq.(\ref{Estar}) remain independent of $B$. This result is a consequence 
of our model, since we have explicitly excluded a field dependence
for these quantities and taken $\rho_n$ to be temperature and 
field independent. Although Volovik \cite{Vol} has shown that
the specific heat has a $B$-dependence in type II materials above the
lower critical field, we have checked that his scaling prediction at 
low temperatures gives only a weak dependence in the range of 
fields $B << H_{c2}$ relevant to our experiment.
Fig. \ref{FTEB} shows the change in 
the electron temperature as a function
of the applied electric field. The rise in
temperature and corresponding decrease in $H_{c2}$ result in decreasing
differential conductivity which leads to the instability. 
Fig. \ref{FTT0} shows the relatively small effect of increasing the
phonon temperature, up to about 40K, on
the final temperature $T^{*}$ reached by the electron gas. 

\section{\label{sec:level3}Energy transfer rate}

\begin{figure}[ht]
\epsfxsize=50mm
\epsffile{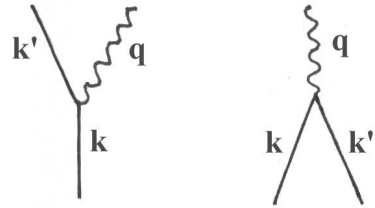}
\caption{Diagrams contributing to the energy transfer rate from the 
quasiparticle gas to the lattice.}
\label{diagrams}
\end{figure}

The total rate at which energy is radiated by the heated quasiparticle gas
to the lattice can be calculated by standard methods \cite{Kea,Well}.
The two contributing processes, phonon emission and quasiparticle
recombination with emission of a phonon, 
are illustrated schematically in Fig. \ref{diagrams}.
In the following calculation, applicable to $d$-wave superconductors, we
assume the Fermi surface to be a cylinder of radius $k_F$ and
height $2\pi/c_0$, $c_0$ being the $c$-axis lattice constant.
The rate for emission of a phonon of momentum ${\bf q}={\bf k} - {\bf k'}$
by a quasiparticle of momentum ${\bf k}$ is
\begin{equation}
w = \frac{V}{(2\pi)^2\hbar}\int\int d^3k' d\omega|
\mathcal{M}_{\bf q}|^2 \delta(\omega - \omega_{\bf q})
\delta(E_{\bf k} - E_{\bf k'} - \hbar\omega).
\end{equation}
This rate can be expressed in terms of the electron-phonon spectral 
function $\alpha^2(\omega)F(\omega)$, defined by
\begin{equation}
\alpha^2(\omega)F(\omega) = \frac{V}{(2\pi)^3\hbar^2}\int\frac{dS'}{v'_F}
|\mathcal{M}_{\bf k - k'}|^2 \delta(\omega - \omega_{\bf k - k'}),
\end{equation}
where $dS'$ is an area element on the Fermi surface and $v'_F$ is the
Fermi velocity.  If the quantity $|\mathcal{M}_{\bf k - k'}|^2
\delta(\omega - \omega_{\bf k - k'})$ is replaced by its average over the 
Fermi surface, 
\begin{equation}
\frac{1}{S}\int dS'|\mathcal{M}_{\bf k - k'}|^2
\delta(\omega - \omega_{\bf k - k'}) = \frac{2\pi c_0 \hbar^2 v_F}{Vk_F}
\alpha^2F(\omega),
\end{equation}
the transition rate becomes 
\begin{equation}
w = \frac{\pi}{3c_0\hbar\gamma}\int d^2k'
\alpha^2F(E_{\bf k} - E_{\bf k'})/\hbar),
\end{equation}
where $d^2k' = k'dk'd\theta'$ and $\theta'$ is the azimuthal angle on the
Fermi cylinder. The total energy transfer rate was obtained by integrating
$w$ over initial quasiparticle energies, and reduces to
\begin{equation}
P_e = \frac{V}{12\pi c_0^2\hbar\gamma}\int d^2k \int d^2k' 
\alpha^2F(E - E')/\hbar)(E - E')g_e(E, E').
\end{equation}
The factor $g_e(E,E')$ contains the occupation factors and coherence factors
\begin{equation}
g_e(E,E') = f(E)(1 - f(E'))
\left((1 - \frac{\Delta\Delta'\cos2\theta\cos2\theta'}
{EE'}\right) 
\end{equation}
for the initial and final states. The second term in the coherence factor
integrates to zero because of the $d$-wave symmetry of the order parameter.
After transforming the momentum integrals into integrals over the quasiparticle
energy and performing the azimuthal integrals, the energy transfer rate 
reduces to
\begin{widetext}
\begin{equation}
P_e = \frac{3V\gamma}{4\pi^3\hbar}\int EdED(E, \Delta)\int E'dE'D(E', \Delta)
F((E-E')/\hbar) (E-E')g_e(E, E').
\end{equation}
\end{widetext}
Here $D(E, \Delta)$ is the $d$-wave density of states
\begin{equation}
\int\frac{d\theta}{\sqrt{E^2 - \Delta^2\cos^2 2\theta}} =
\left\{
\begin{array}{cc}
\frac{4}{\Delta}K(\frac{E^2}{\Delta^2}), & E < \Delta \\
\frac{4}{E}K(\frac{\Delta^2}{E^2}), & E > \Delta, 
\end{array}
\right.
\label{DOS}
\end{equation}
where $K$ denotes the elliptic function of the first kind.
In this way of calculating, there is some averaging over the Fermi surface,
but the characteristic $d$-wave density of states with its logarithmic 
singularity at $E = \Delta$ has been retained.

The spectral function $\alpha^2F(\omega)$ is assumed to be of the 
form $b\omega^2$ appropriate for acoustic phonons. Although optical phonons
are present in high temperature superconductors, they are assumed to make
only a negligible contribution to the thermal conductivity responsible for
carrying energy from the heated electrons to the bath. The spectral function 
is cut off at the Debye frequency $\omega_D$. The constant $b$ measures the
strength of the electron-phonon interaction. This is usually expressed
through the electron-phonon coupling constant $\lambda$, defined as the
integral of $2\alpha^2F(\omega)/\omega$ over frequency, which has a value
of order unity in most superconductors. Since we are only considering 
acoustic phonons, this provides an upper limit for $b$ of about
$10^{-3} {\rm meV}^{-2}$. This is consistent with the magnitude of the
electron-phonon matrix element quoted in Ref.[\onlinecite {AM}], which gives
$b \approx 6\times10^{-4}$. It is also in the same range as the values
extracted from neutron scattering data by Kaplan {\em et al.}\cite{Kea} for
low temperature superconductors. The deformation potential approximation
as cited in Ref.[\onlinecite{Well}],
on the other hand, gives a value of the order of $3\times10^{-7} 
{\rm meV}^{-2}$. The choice that gives the best peak values for the
peak current and for $E^*$ is $5.2\times10^{-6}{\rm meV}^{-2}$, more in
line with the latter value. The curves shown have taken this best fit 
value for $b$.

Introducing the dimensionless variables $x=E/T, y=E'/T$, and $z=\Delta/T$,
the energy transfer rate takes the form
\begin{equation}
P_e = GVT^5\Phi_e,
\end{equation}
with
\begin{equation}
G=\frac{3\gamma b}{4\pi^3\hbar^3},
\end{equation}
and the dimensionless integral $\Phi_e$ given by
\begin{equation}
\Phi_e = \int_0^{\infty}dx\int_0^x\frac{D(x,z)D(y,z)xy(x-y)^3}{
e^x + e^{x-y}+e^{-y}+1}.
\end{equation}
$D(x,y)$ is the dimensionless form of the density of states Eq. [\ref{DOS}].
The quasiparticle recombination process gives an expression of the same 
form with $\Phi_e$ replaced by 
\begin{equation}
\Phi_r = \int_0^{d}dx\int_0^{d-x}dy\frac{D(x,z)D(y,z)xy(x+y)^3}{
e^x + e^{x+y}+e^y+1}.
\end{equation}
The value of the specific heat constant $\gamma$ is taken from measurements
of W. C. Lee, et. al.\cite{Lee} as 0.0331 meV$^{-1}$ nm$^{-3}$, giving
$G = 1.257 \times 10^6$ 
meV$^{-1}$ nm$^{-3}$ s$^{-1}$ when $b$ is normalized to give $\lambda = 1.$

\begin{figure}[hb]
\epsfxsize=70mm
\epsffile{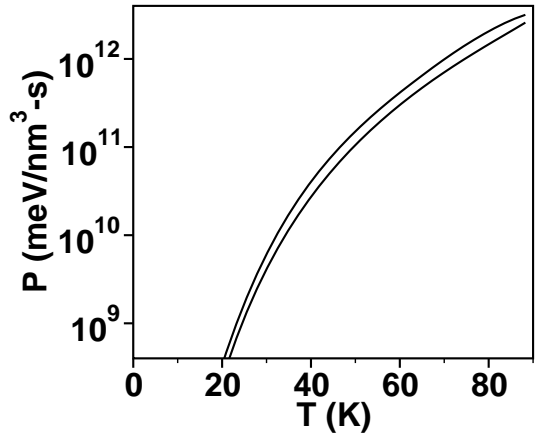}
\caption{Calculated energy transfer rate in $\textrm{meV/nm}^3$-s of 
electron gas at temperature $T$
to a zero temperature lattice. The lower curve represents the phonon emission 
process and the upper curve represents the quasiparticle annihilation process.}
\label{FPER}
\end{figure}
\begin{figure}[htp]
\epsfxsize=70mm
\epsffile{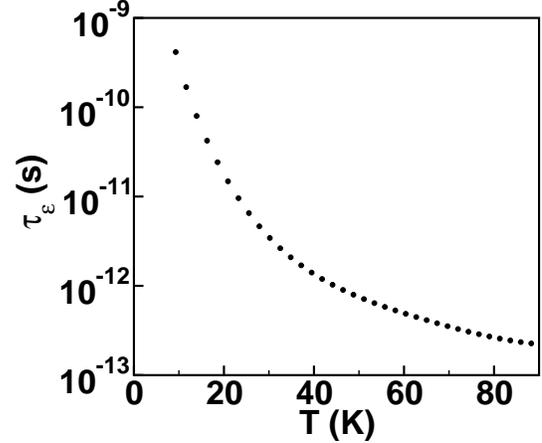}
\caption{Energy relaxation time \teps as a function of temperature.}
\label{Ftau}
\end{figure}

 
Fig. \ref{FPER} shows the calculated energy transfer rates for emission and
recombination. Fig. \ref{Ftau} shows the energy relaxation time \teps found 
by equating the total
transfer rate from both processes to the last term in
the rate equation (Eq.~\ref{dudt}). 
 
We note the following properties of the energy transfer rate:

1.) The rate of energy 
transfer from the electrons to the lattice
at any given temperature is equal to the rate of transfer 
from the lattice to the electrons at the same temperature. 
Indeed, the rate from phonon emission balances the rate from
absorption and the rate from quasiparticle recombination balances
the rate from pair creation. These results can be demonstrated 
in the deformation
potential approximation, where the matrix element for phonon
emission and absorption depends only on the phonon energy $\nu$.
For example, the rates for emission and absorption and for quasiparticle
recombination and creation
can be written
\begin{widetext}
\begin{eqnarray}
P_e & = & G\int (n(\nu) + 1)
f_e(E,E - \nu)N(E)N(\nu - E)\nu^3 dEd\nu \nonumber \\
P_a & = & G\int n(\nu) 
f_a(E,E + \nu)N(E)N(E + \nu)\nu^3 dEd\nu, \nonumber \\
P_r & = & G\int (n(\nu) + 1)
f_r(E,\nu - E)N(E)N(\nu - E)\nu^3 dEd\nu \nonumber \\
P_c & = & G\int n(\nu) 
f_c(E,\nu - E)N(E)N(\nu - E) \nu^3dEd\nu, \nonumber 
\end{eqnarray}
\end{widetext}
where $n(\nu)$ is the phonon occupation number at the given
temperature and $N(E)$ is the quasiparticle density of states, and $G$
has the value $3Vb\gamma/4\pi^3 \hbar^3$. 
The equality of the rates is evident upon substituting
the explicit forms of the Fermi and Bose distribution functions.
In the same manner, the emission and absorption rates
are identical after the same substitutions and the change of variable
$E^{'} = E + \nu$ in $P_a$.

2.) The differences between the emission and absorption
rates and between the pair recombination and creation rates
have only a weak dependence on the lattice temperature as long as the 
electron temperature is near $T_c$ and the lattice temperature is low,
say $T_p \leq T_c/2$. This conclusion is based on values for $T_c$ 
(7.75 meV) and $\Delta$ (19 meV) for YBCO. Wellstood, Urbina, and Clarke, 
\cite{Well}, assert that
the difference between the emission rate and the absorption rate
for a normal metal is equal to the difference between the rate electrons
radiate to a zero temperature lattice and rate phonons radiate to a zero 
temperature electron gas.
This result is only approximately valid in the gas of quasiparticles.
The differences can be calculated from the previous pairs of equations
by taking $n(\nu)$ to be the phonon distribution function at $T_p$. 
The difference between emission and absorption rates, for example, is:
\begin{displaymath}
P_e - P_a = G\int_{0}^{\infty}dE\int_0^{E-\Delta}d\nu 
f_e(E, E - \nu)\frac{e^{\nu/T_p} - e^{\nu/T'}}{e^{\nu/T_p} - 1}.
\end{displaymath} 
The dependence on $T_p$ is contained in the last factor. For $\nu$ of the 
order of $\Delta$, $T'$ of the order $T_c$, and $T_0$ in the range 
zero to $T_c/2$, this factor
only varies from 1.0 to 0.915, showing therefore a weak dependence of the 
difference on $T_p$. In an earlier calculation assuming $s$-wave symmetry
of the order parameter, the differences were calculated explicitly
for $T=0.8\;T_c$ and $T_p$ ranging from $0.1\;T_c$ to $0.5\;T_c$. The 
difference varies less than 10\% for emission and absorption and less than
1\% for the dominant creation and recombination. In view of these results,
we ignore the dependence on $T_p$ and calculate \teps\bb on the basis
of the radiation rate to a zero temperature lattice.
A very similar argument applies to the difference
between the pair recombination and pair creation rates. 

3.) Quasiparticle emission and absorbtion can only satisfy the energy 
and momentum conservation laws if the the quasiparticle velocity
$v_F \partial E / \partial \epsilon$ before emission or after 
absorbtion is greater than the sound velocity.
This \v{C}erenkov condition should be taken into account 
in the averaging near the
Fermi surface that enters into calculation of the electron-phonon
spectral function $\alpha^2 F$. In the integrals over quasiparticle
energy above, the lower limit should be the energy $E_c$ at which
quasiparticles reach the sound velocity rather than 0.  
The correction is of the order of the square of the ratio of the 
sound velocity to the Fermi velocity. Since $s/v_F << 1$ for all
superconductors ($s/v_F \approx 1.5 \times 10^{-2}$ in YBCO), 
the correction can be safely ignored.

\section{\label{sec:level4}Phonon lifetime effects}

In the above discussion, we have not distinguished the phonon temperature
and the bath temperature. We now consider corrections arising from
a more general treatment of the phonon distribution.
The standard estimate of phonon mean free path for normal metals
$\hbar v_F/kT$ gives 17 nm when the Fermi velocity is taken to be
$2\times10^5$ m/s, which is shorter than the 100 nm thickness of
the experimental films. The estimate of Kaplan, Chu, Langenberg, Chang, 
Jafarey, and Scalapino \cite{Kea} 
of quasiparticle and phonon lifetimes in an {\em s}-wave superconductor
below the critical temperature
gives a frequency- and temperature-dependent numerical factor 
of order unity multiplied by the characteristic time
\begin{equation}
\tau^{\textit{ph}}_0 = \frac{\hbar N \langle \alpha^2 \rangle_{\textit{av}}}
{4 \pi^2 N(0)}\Delta(0), \label{tau0}
\end{equation}
where $N$ is the ion number density, $\langle \alpha^2 \rangle_{\textit{av}}$
is the average electron-phonon coupling constant, $N(0)$ is the single-spin
electronic density of states at the Fermi surface, and $\Delta(0)$ is the
zero-temperature gap. Taking \cite{nist} the values $N$ = 13 per unit cell, 
$\langle \alpha^2 \rangle_{\textit{av}} = 5$ meV, $N(0)$ calculated from
the free-electron theory with $v_F$ having the value quoted above, and
$\Delta(0)$ = 19 meV, and converting the lifetime to a mean free path
using the longitudinal sound velocity $4.2\times10^3$ m/s yields a 
path of the order of $10^3$ nm, an order of magnitude larger than 
the thickness of the experimental film. 

These estimates indicate that we are dealing with a case in which 
the phonon mean free path could be comparable to the thickness of the sample.
To deal with the general case, we follow Bezuglij and Shklovskij \cite{BjSy}, 
writing the kinetic equation for the phonon 
distribution function $n({\bf q}, z)$ as
\begin{equation}
s_z\frac{\partial n({\bf q}, z)}{\partial z} = -\frac{n({\bf q},z) - n(T)}
{\tau_{\textit{ph}}}, \label{KinE} 
\end{equation}
where $s_z$ is the component of the sound velocity perpendicular to 
the plane of the film and $n(T)$ is the thermal phonon distribution at the
electron temperature. If phonons are reflected at the free
surface of the film and transmitted with average coefficient $\alpha$ at
the substrate interface, it is found that the phonon distribution 
function is a linear combination of two thermal distributions, one at
the bath temperature and one at the electron temperature. The coefficients
in the linear combination depend on the position within the film and
on the direction of propagation of the phonons:
\begin{equation}
n = A(z, \theta)n(T) + B(z, \theta)n(T_0). \label{dist} 
\end{equation} 
The integrand in the expression for the energy transfer rate 
from quasiparticle gas to lattice contains the factor $n + 1$, while
that for the reverse rate contains a factor $n$. If Eq.(\ref{dist})
is substituted into these rates and account is taken of the 
condition for equilibrium between the lattice and the gas, the resulting
rate contains a term with the factor $1 - A$ and a term with the 
factor $Bn(T_0)$. For purposes of estimating the correction for
finite phonon lifetime, we neglect the term proportional to $n(T_0)$
compared to the $1 - A$ term on the ground that the 
phonon number is small at a temperature $T_0$ which is much 
smaller than the Debye temperature.
An estimate of the remaining term can
be obtained by replacing $1 - A$ in the integral for the rate by its 
average value over the 
thickness of the film and over the directions of propagation of the phonon.
The remaining integral is the one we evaluated in the previous section.

The resulting explicit expression for $A$ is:
\begin{equation}
A = 1 - \frac{\alpha}{1 - (1 - \alpha)e^{-2d/l_z}}\left\{\begin{array}{lcl}
e^{-z/l_z} & , & q_z > 0 \\e^{-(2d - z)/l_z} & , & q_z < 0 \end{array}\right\}.
 \label{A}
\end{equation}
Eq.~\ref{dist} with \ref{A} reflects the gradual change of the distribution 
from a
nearly thermal distribution at the bath temperature at the substrate
interface $z = 0$ to an electron temperature 
thermal distribution over the distance of a phonon mean free path.
The transmission probability $\alpha$ can be determined 
in principle \cite{Sy} from the measured 
value of the thermal resistance of the film-substrate interface, 
defined as the ratio of $\Delta T$ at the interface to the product of
the power dissipated per unit volume and the thickness of the film.
The measured value\cite{Nah,mplb}  for YBCO is about 
$1 \times 10^{-3}$ Kcm$^2$/W.
The determination of $\alpha$ is affected by uncertainties due to the 
averaging and due to the temperature variation of the 
thermal resistance.  
A literal application of Eq.[14] of Ref.[\onlinecite{Sy}] produces 
the average value 0.184
when $d$ is of the order of or larger than $l$. When $d \ll l$, 
Shklovskii shows that the effective $\alpha$ is $2d/l$, which is
0.2 for the longest estimate of phonon mean free path above.
We therefore accept 0.2 as a reasonable value. 
Sensitivity of the value of $1 - A$ to $d/l$ and $\alpha$ are shown in
Table \ref{tab1}.
The longest estimate of phonon mean free path with the best estimate
of the transmission coefficient indicate that the energy transfer rate
will be multiplied by a factor of 0.325 due to phonon lifetime effects. 

\begin{table}[ht]
\begin{tabular}{|c|c|c|c|}
\hline
\hspace{2mm}$d/l$\hspace{2mm} & \hspace{2mm}$\alpha = 0.8$\hspace{2mm} 
& \hspace{2mm}$\alpha = 0.5$\hspace{2mm} 
& \hspace{2mm}$\alpha = 0.2$\hspace{2mm} \\
\hline
6.0 & 0.0333 & 0.0208 & 0.00833 \\
\hline
1.0 & 0.190 & 0.121 & 0.0493 \\
\hline
0.1 & 0.680 & 0.553 & 0.325 \\
\hline
\end{tabular} 
\caption{Calculated values of the phonon lifetime factor 
$1 - A$ for three values of
the ratio of thickness $d$ to phonon mean free path $l$ and three 
values of the average transmission coefficient $\alpha$.} 
\label{tab1}
\end{table}

\section{\label{sec:level5}Conclusion}

At temperatures well below \tcy, 
high electric fields and current densities can produce an instability that 
can be accounted for by 
a hot electron gas model in which the electronic temperature is elevated 
due to dissipation. 
The calculations presented here provide a quantitative justification
for this scenario by showing that the 
temperature variation of the energy transfer rate between the
lattice and the electrons is consistent with the position of the
instability observed in YBCO films.  
They show also that if the phonon mean free path is not too small compared to 
the film thickness, the necessary temperature difference between electrons
and lattice can be maintained.  

\section{\label{sec:level6}Acknowledgements}

We thank B. I. Ivlev for important suggestions, and M. Geller, R. P.
Huebener, and R. P. Prozorov for helpful discussions and criticisms.
This work was supported by the U. S. Department of
Energy through grant number DE-FG02-99ER45763.

\end{document}